# Effect of the volume of the drainage system on the measurement of undrained thermo-poro-elastic parameters

Siavash Ghabezloo[*], Jean Sulem

*Université Paris-Est, UR Navier, CERMES, Ecole des Ponts ParisTech, Marne la Vallée, France*

## Abstract

For evaluation of the undrained thermo-poro-elastic properties of saturated porous materials in conventional triaxial cells, it is important to take into account the effect of the dead volume of the drainage system. The compressibility and the thermal expansion of the drainage system along with the dead volume of the fluid filling this system, influence the measured pore pressure and volumetric strain during an undrained thermal or mechanical loading in a triaxial cell. The correction methods previously presented by Wissa (1969), Bishop (1976) and Ghabezloo and Sulem (2009) only permit to correct the measured pore pressures during an undrained isotropic compression test or an undrained heating test. An extension of these methods is presented in this paper to correct also the measured volumetric strain and consequently the measured undrained bulk compressibility and undrained thermal expansion coefficients during these tests. Two examples of application of the proposed correction method are presented on the results of an undrained isotropic compression test and an undrained heating test performed on a fluid-saturated granular rock. A parametric study has demonstrated that the porosity and the drained compressibility of the tested material, and the ratio of the volume of the drainage system to the one of the tested sample are the key parameters which influence the most the error induced on the measurements by the drainage system.

[*]Corresponding Author: Siavash Ghabezloo, CERMES, Ecole des Ponts ParisTech, 6-8 avenue Blaise Pascal, Cité Descartes, 77455 Champs-sur-Marne, Marne la Vallée cedex 2, France, Email: ghabezlo@cermes.enpc.fr

# 1 Introduction

The undrained condition is defined theoretically as a condition in which there is no change in the fluid mass of the porous material. For performing an undrained test in the laboratory, this condition cannot be achieved just by closing the valves of the drainage system as it is done classically in a conventional triaxial system (Figure (1)). In a triaxial cell, the tested sample is connected to the drainage system of the cell and also to the pore pressure transducer. As the drainage system has a non-zero volume filled with water, it experiences volume changes due to its compressibility and its thermal expansivity. The variations of the volume of the drainage system and of the fluid filling the drainage system induce a fluid flow into or out of the sample to achieve pressure equilibrium between the sample and the drainage system. This fluid mass exchanged between the sample and the drainage system modifies the measured pore pressure and consequently the measured strains during the test.

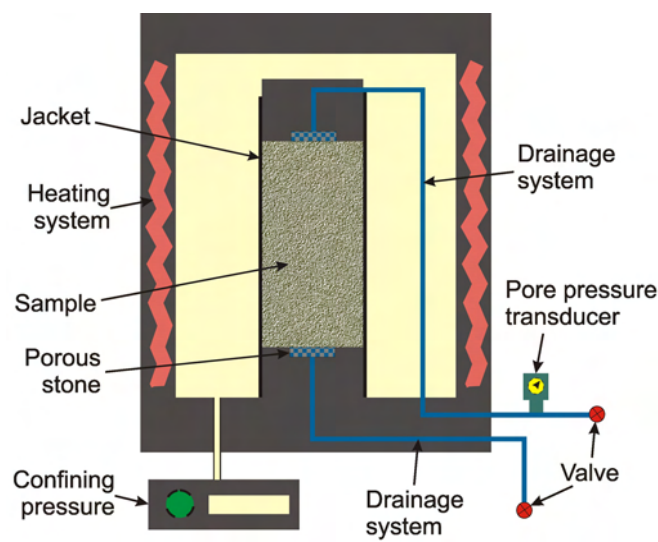

**Figure 1- Schematic view of a conventional triaxial cell**

Wissa [1] was the first who studied this problem for a mechanical undrained loading. He presented an expression for the measured pore pressure increase as a function of the compressibilities of pore-water, soil skeleton, pore-water lines and pressure measurement system, but the compressibility of the solid phase was not considered in his work. Bishop [2] presented an extension of the work of Wissa [1] taking into account the compressibility of the solid grains. The proposed method was first used by Mesri *et al.* [3] to correct the measured pore pressure in undrained isotropic compression tests. Ghabezloo and Sulem [4] have presented an extension to the work of Bishop [2] to correct the pore pressure measured during undrained heating and cooling tests by taking into account the thermal expansion of the drainage system, the inhomogeneous temperature distribution in the drainage system and also the thermal expansion of the fluid filling the drainage system. The proposed method was applied to the results of undrained heating tests performed on a Rothbach sandstone [4] and on a hardened cement paste [5]. The correction methods presented by Wissa [1], Bishop [2] and Ghabezloo and Sulem [4] deal with the pore pressure measured during an undrained mechanical or thermal loading. But it is obvious that the measured strain during an undrained test is also influenced in the same way by the effect of the deformation of the drainage system and by the fluid mass exchange between the sample and the drainage system. In this paper the corrections presented by Bishop [2] and Ghabezloo and Sulem [4] are extended for the correction of the measured strains and consequently the correction of the measured Skempton coefficient and the measured undrained thermal expansion coefficient. Moreover, the poroelastic framework used by Bishop [2] and Ghabezloo and Sulem [4] is suited for an ideal porous material composed of one single solid phase. In this paper, this framework is extended to a more general case of a porous material which can be heterogeneous and anisotropic at the microscale. An additional improvement as compared to the previous correction methods is also brought by taking into account the dependence of the dead volume of the drainage system on the applied confining pressure.

## 2 Poroelastic framework

A fluid saturated porous material can be seen as a mixture of two phases: a fluid phase and a solid phase which may be made up of several constituents. The (Eulerian) porosity $\phi$ is defined as the ratio of the volume of the porous space $V_\phi$ to total volume $V$ in the actual (deformed) state:

$$\phi = \frac{V_\phi}{V} \tag{1}$$

We consider a saturated sample under an isotropic state of stress $\sigma$ (positive in compression) and we define the differential pressure $\sigma_d$ (i.e. Terzaghi effective stress) as the difference between the confining pressure $\sigma$ and the pore pressure $p_f$.

$$\sigma_d = \sigma - p_f \tag{2}$$

The variations of the total volume $V$ and of the pore volume $V_\phi$ are given in the following expressions as a function of variations of three independent variables, $\sigma_d$ and $p_f$ and the temperature $T$:

$$\frac{dV}{V} = -c_d d\sigma_d - c_s dp_f + \alpha_d dT \tag{3}$$

$$\frac{dV_\phi}{V_\phi} = -c_p d\sigma_d - c_\phi dp_f + \alpha_\phi dT \tag{4}$$

where $c_d$, $c_s$, $c_p$ and $c_\phi$ are four elastic compressibility coefficients, and $\alpha_d$ and $\alpha_\phi$ are two thermal expansion coefficients defined below:

$$c_d = -\frac{1}{V}\left(\frac{\partial V}{\partial \sigma_d}\right)_{p_f,T} \quad , \quad c_p = -\frac{1}{V_\phi}\left(\frac{\partial V_\phi}{\partial \sigma_d}\right)_{p_f,T} \tag{5}$$

$$c_s = -\frac{1}{V}\left(\frac{\partial V}{\partial p_f}\right)_{\sigma_d,T} \quad , \quad c_\phi = -\frac{1}{V_\phi}\left(\frac{\partial V_\phi}{\partial p_f}\right)_{\sigma_d,T} \tag{6}$$

$$\alpha_d = \frac{1}{V}\left(\frac{\partial V}{\partial T}\right)_{p_f,\sigma_d} \quad , \quad \alpha_\phi = \frac{1}{V_\phi}\left(\frac{\partial V_\phi}{\partial T}\right)_{p_f,\sigma_d} \tag{7}$$

Equation (5) corresponds to an isothermal drained isotropic compression test in which the temperature and the pore pressure are kept constant inside the sample while the confining pressure is varied. The variations of the total volume of the sample $V$ and of the volume of the pore space $V_\phi$ with respect to the applied confining pressure give the (isothermal) drained bulk modulus $K_d = 1/c_d$ and the modulus $K_p = 1/c_p$. Equation (6) corresponds to an isothermal unjacketed compression test, in which equal increments of confining pressure and pore pressure are simultaneously applied to the sample. The differential pressure $\sigma_d$ in this case remains constant and the measured volumetric strain with the applied pressure gives the unjacketed modulus $K_s = 1/c_s$. The variation of the pore volume of the sample in this test, evaluated from the quantity of fluid exchanged between the sample and the pore pressure/volume controller when applying equal increments of confining pressure and pore pressure could in principle give the modulus $K_\phi = 1/c_\phi$. However, experimental evaluation of this parameter is very difficult because the volume of the exchanged fluid has to be corrected for the effect of fluid compressibility and also for the effect of the deformations of the pore pressure/volume controller and of the drainage system. However, the variation of the pore volume of the sample is very small as compared to the volume of the fluid exchanged between the sample and the pore pressure/volume controller, and thus the correction might not be relevant. In the case of a porous material which is composed of two or more solids and therefore is heterogeneous at the micro-scale, the unjacketed compressibility $c_s$ is some weighted average of the compressibilities of solid constituents [6]. What this average should be is generally unknown, however, in Ghabezloo and Sulem [4] the unjacketed modulus of Rothbach sandstone was evaluated using Hill's [7] average formula and was in good accordance with the experimentally evaluated modulus. The compressibility $c_\phi$ for such a material has a complicated dependence on the material properties. Generally it is not bounded by the elastic compressibilities of the solid components and can even have a negative sign if the compressibilities of the individual solid components are greatly different one from another [8,9].

Equation (7) corresponds to a drained heating test in which the pore pressure and the confining pressure are controlled to remain constant while the thermal loading is applied. The variations of the total volume of the sample $V$ and of the volume of the pore space $V_\phi$ with respect to the temperature give the volumetric drained thermal expansion coefficient $\alpha_d$ and pore volume thermal expansion coefficient $\alpha_\phi$. Like for $c_\phi$, the experimental evaluation of $\alpha_\phi$ is very difficult because the volume of the exchanged fluid has to be corrected for the effect of the thermal expansion of the fluid and also for the effect of the thermal deformations of the pore pressure generator and of the drainage system in order to access to the variation of the pore volume of the sample. In the case of a micro-homogeneous and micro-isotropic porous material, $c_s = c_\phi = c_m$ and $\alpha_d = \alpha_\phi = \alpha_m$, where $c_m$ and $\alpha_m$ are respectively the compressibility and the thermal expansion coefficient of the single solid constituent of the porous material. For such a material, there is no change of porosity during an unjacketed compression test or a drained heating test.

Using Betti's reciprocal theorem the following relation can be obtained between the elastic moduli [10,11]:

$$c_p = \frac{c_d - c_s}{\phi} \qquad (8)$$

Using equation (8), the number of the required parameters to characterize the volumetric thermo-poro-elastic behaviour of a porous material is reduced to five, among which the experimental evaluation of $c_\phi$ and $\alpha_\phi$ is very difficult as mentioned above. Nevertheless, we can access indirectly to these parameters by performing undrained tests.

The undrained condition is defined as a condition in which the mass of the fluid phase is constant ( $dm_f = 0$ ). Under this condition we choose three different independent variables: the total stress $\sigma$, the fluid mass $m_f$, and the temperature $T$. Writing the expression of the variation of the total volume $V$ and the pore pressure $p_f$ with the variations of the total stress $\sigma$ and the temperature $T$, we can define four new parameters to describe the response of the porous material in undrained condition:

$$B = \left(\frac{\partial p_f}{\partial \sigma}\right)_{m_f,T} \quad , \quad c_u = -\frac{1}{V}\left(\frac{\partial V}{\partial \sigma}\right)_{m_f,T} \qquad (9)$$

$$\Lambda = \left(\frac{\partial p_f}{\partial T}\right)_{m_f,\sigma} \quad , \quad \alpha_u = \frac{1}{V}\left(\frac{\partial V}{\partial T}\right)_{m_f,\sigma} \qquad (10)$$

The parameter $c_u$ is the undrained bulk compressibility, $B$ is the so-called Skempton [12] coefficient, $\alpha_u$ is the undrained volumetric thermal expansion coefficient and $\Lambda$ is the thermal pressurization coefficient. For a saturated porous material the fluid mass is given by $m_f = V_\phi \rho_f$, where $\rho_f$ is the pore-fluid density. Fluid mass conservation under undrained condition ($dm_f = 0$) leads to the following expression for variation of the volume of the fluid:

$$\frac{dV_\phi}{V_\phi} = -c_f dp_f + \alpha_f dT \tag{11}$$

where $c_f$ and $\alpha_f$ are respectively the pore-fluid compressibility and thermal expansion coefficient. Replacing equation (11) in equation (4) and using equations (2) and (8) the following expressions are obtained for the Skempton coefficient $B$ and the thermal pressurization coefficient $\Lambda$:

$$B = \frac{(c_d - c_s)}{(c_d - c_s) + \phi(c_f - c_\phi)} \tag{12}$$

$$\Lambda = \frac{\phi(\alpha_f - \alpha_\phi)}{(c_d - c_s) + \phi(c_f - c_\phi)} \tag{13}$$

Gassmann [13] was the first who presented an equivalent from of equation (12) by supposing $c_\phi = c_s$, but the current form of this equation was first presented by Brown and Korringa [10]. The expressions of the thermal pressurization coefficient, equivalent to equation (13), were presented by Palciauskas and Domenico [14] and McTigue [15]. The variation of the total volume in undrained condition is given by the undrained bulk compressibility $c_u$ and the undrained thermal expansion coefficient $\alpha_u$. Replacing $dV/V = -c_u d\sigma + \alpha_u dT$, $d\sigma_d = d\sigma - dp_f$ and $dp_f = Bd\sigma + \Lambda dT$ in equation (3), the following relationships are found for $c_u$ and $\alpha_u$:

$$c_u = c_d - B(c_d - c_s) \tag{14}$$

$$\alpha_u = \alpha_d + \Lambda(c_d - c_s) \tag{15}$$

Equations (12) to (15) can be used for an indirect evaluation of the parameters $c_\phi$ and $\alpha_\phi$ as functions of the other poroelastic parameters which are easier to evaluate experimentally. See for example Ghabezloo *et al.* [16] for an indirect evaluation of the compressibility $c_\phi$.

# 3   Correction of the effect of drainage system

In a triaxial cell the tested sample is connected to the drainage system of the cell and the undrained condition is achieved by closing the valves of this system (Figure (1)). Consequently, the condition $dm_f = 0$ is applied to the total volume of the fluid which fills the pore volume of the sample and also the drainage system:

$$m_f = V_\phi \rho_f + V_L \rho_{fL} \tag{16}$$

where $V_L$ is the volume of the drainage system and $\rho_{fL}$ is the density of the fluid in the drainage system. As the sample and the drainage system may have different temperatures and considering that the fluid density varies with temperature, different densities are considered for the pore-fluid of the sample and for the fluid filling the drainage system. The variation of volume of the drainage system can be written in the following form:

$$\frac{dV_L}{V_L} = c_L dp_f + \alpha_L dT_L - \kappa_L d\sigma \tag{17}$$

where $dT_L$ is the equivalent temperature change in the drainage system, $c_L$ and $\kappa_L$ are isothermal compressibilities and $\alpha_L$ is the thermal expansion coefficient of the drainage system defined as:

$$c_L = \frac{1}{V_L}\left(\frac{\partial V_L}{\partial p_f}\right)_{T_L,\sigma} \tag{18}$$

$$\alpha_L = \frac{1}{V_L}\left(\frac{\partial V_L}{\partial T_L}\right)_{p_f,\sigma} \tag{19}$$

$$\kappa_L = -\frac{1}{V_L}\left(\frac{\partial V_L}{\partial \sigma}\right)_{p_f,T_L} \tag{20}$$

The parameter $c_L$ is equivalent to $\left(c_L + c_M\right)/V_L$ in Wissa [1] and Bishop [2]. Based on the data provided by these authors, some typical values of $c_L$ can be evaluated which vary between 0.008 GPa$^{-1}$ and 0.16 GPa$^{-1}$. It should be mentioned that the parameters $c_L$ and $\alpha_L$ defined in equations (18) and (19) are equivalent respectively to $c_L/V_L$ and $\alpha_L/V_L$ in Ghabezloo and Sulem [4].

In most triaxial devices, the drainage system can be separated into two parts: one situated inside the triaxial cell and the other one situated outside the cell. In the part inside the cell, one can assume that the temperature change $dT$ is identical to the one of the sample; in the other part situated outside the cell, the temperature change is smaller than $dT$ and varies along the drainage lines. We define an equivalent homogeneous temperature change $dT_L$ such that the volume change of the entire drainage system caused by $dT_L$ is equal to the volume change induced by the true non-homogeneous temperature field. The temperature ratio $\beta$ is an additional parameter which is defined below and evaluated on a calibration test as explained further:

$$\beta = \frac{dT_L}{dT} \tag{21}$$

By writing the undrained condition $dm_f = 0$, using equation (17) and taking into account the variations of the fluid density with pore pressure and temperature changes, the following expression is obtained for the variations of the pore volume of the tested sample:

$$\frac{dV_\phi}{V_\phi} = -c_f dp_f + \alpha_f dT + \frac{V_L}{V_\phi}\frac{\rho_{fL}}{\rho_f}\left(-c_{fL}dp_f - c_L dp_f + \alpha_{fL}\beta dT - \alpha_L \beta dT + \kappa_L d\sigma\right) \tag{22}$$

As the thermal expansion coefficient and the compressibility of water both vary with temperature, the parameters used for the water in the drainage system, $\alpha_{fL}$ and $c_{fL}$, are different from the parameters used for the pore fluid of the porous material. Replacing equation (22) in equation (4) and using equations (2) and (8), the expressions of the measured Skempton coefficient and thermal pressurization coefficient, $B^{mes}$ and $\Lambda^{mes}$ are obtained.

$$B^{mes} = \frac{\left(c_d - c_s\right) + \dfrac{V_L}{V}\dfrac{\rho_{fL}}{\rho_f}\kappa_L}{\left(c_d - c_s\right) + \phi\left(c_f - c_\phi\right) + \dfrac{V_L}{V}\dfrac{\rho_{fL}}{\rho_f}\left(c_{fL} + c_L\right)} \tag{23}$$

$$\Lambda^{mes} = \frac{\phi\left(\alpha_f - \alpha_\phi\right) + \beta\dfrac{V_L}{V}\dfrac{\rho_{fL}}{\rho_f}\left(\alpha_{fL} - \alpha_L\right)}{\left(c_d - c_s\right) + \phi\left(c_f - c_\phi\right) + \dfrac{V_L}{V}\dfrac{\rho_{fL}}{\rho_f}\left(c_{fL} + c_L\right)} \tag{24}$$

The comparison of the equations (23) and (24) respectively with equations (12) and (13) shows the effect of the drainage system of the triaxial cell on the measured coefficients. Using equations (23), (12) and (14) the expressions of the corrected Skempton coefficient $B^{cor}$ and the corrected undrained bulk compressibility $c_u^{cor}$ are found:

$$B^{cor} = \frac{B^{mes}}{1 + \dfrac{V_L \rho_{fL}}{V \rho_f \left(c_d - c_s\right)} \left[\kappa_L - B^{mes}\left(c_{fL} + c_L\right)\right]} \qquad (25)$$

$$c_u^{cor} = c_d - \frac{c_d - c_u^{mes}}{1 + \dfrac{V_L \rho_{fL}}{V \rho_f \left(c_d - c_s\right)} \left[\kappa_L - \dfrac{c_d - c_u^{mes}}{c_d - c_s}\left(c_{fL} + c_L\right)\right]} \qquad (26)$$

Equation (25) is similar to the one presented by Bishop [2], but differs in two points. The first one is that Bishop did not account for the influence of the confining pressure on the dead volume of the drainage system. This effect appears in equation (25) through the parameter $\kappa_L$. The second one is that Bishop assumed equal densities and compressibilities for the sample's pore fluid and the fluid in the drainage system, which is a correct assumption for an isothermal undrained test performed at ambient temperature. The same assumption is also made by Ghabezloo and Sulem [4]. For an isothermal undrained test performed at an elevated temperature, the situation depends on the heating system of the triaxial cell. If the heating system is such that the temperature change is uniform in the sample and in the drainage system, the assumption of similar properties for the sample's pore fluid and the fluid in the drainage system can be used. Otherwise, as shown in Figure (1), the sample temperature is different from the average temperature in the drainage system so that different properties should be considered for the sample's pore fluid and the fluid in the drainage system.

The correction method proposed here in equations (25) and (26) is applied directly on the results of the test, but it is restricted to an elastic response of the sample and of the drainage system. It differs from the method proposed by Lockner and Stanchits [17] who have modified the procedure of the test itself by imposing a computer-generated virtual 'no-flow boundary condition' at the sample-endplug interface to insure that no volume change occurs in the drainage system.

Using equations (24), (13) and (15) the following expressions are obtained for the corrected thermal pressurization coefficient $\Lambda^{cor}$ and the corrected undrained thermal expansion coefficient $\alpha_u^{cor}$:

$$\Lambda^{cor} = \frac{\Lambda^{mes}}{1 + \dfrac{V_L \rho_{fL}}{V \rho_f \phi\left(\alpha_f - \alpha_\phi\right)} \left[\beta\left(\alpha_{fL} - \alpha_L\right) - \Lambda^{mes}\left(c_{fL} + c_L\right)\right]} \qquad (27)$$

$$\alpha_u^{cor} = \alpha_d + \frac{\alpha_u^{mes} - \alpha_d}{1 + \dfrac{V_L \rho_{fL}}{V \rho_f \phi\left(\alpha_f - \alpha_\phi\right)} \left[\beta\left(\alpha_{fL} - \alpha_L\right) - \left(\alpha_u^{mes} - \alpha_d\right)\dfrac{c_{fL} + c_L}{c_d - c_s}\right]} \qquad (28)$$

## 4 Experimental setting

The triaxial cell used in this study can sustain a confining pressure up to 60MPa. It contains a system of hydraulic self-compensated piston. The loading piston is then equilibrated during the confining pressure build up and directly applies the deviatoric stress. The axial and radial strains are measured directly on the sample inside the cell with two axial transducers and four radial ones of LVDT type. The confining pressure is applied by a servo controlled high pressure generator. Hydraulic oil is used as confining fluid. The pore pressure is applied by another servo-controlled pressure generator with possible control of pore volume or pore pressure.

The heating system consists of a heating belt around the cell which can apply a temperature change with a given rate and regulate the temperature, and a thermocouple which measures the temperature of the heater. In order to limit the temperature loss, an insulation layer is inserted between the heater element and the external wall of the cell. A second insulation element is also installed beneath the cell. The heating system heats the confining oil and the sample is heated consequently. Therefore there is a discrepancy between the temperature of the heating element in the exterior part of the cell and that of the sample. In order to control the temperature in the centre of the cell, a second thermocouple is placed at the vicinity of sample. The temperature given by this transducer is considered as the sample temperature in the analysis of the test results. A schematic view of this triaxial cell is presented in Ghabezloo and Sulem [4] and Sulem and Ouffroukh [18].

# 5   Calibration of the correction parameters

The drainage system is composed of all the parts of the system which are connected to the pore volume of the sample and filled with the fluid, including pipes, pore pressure transducers, porous stones. The volume of fluid in the drainage system $V_L$, can be measured directly or evaluated by using the geometrical dimensions of the drainage system. For the triaxial cell used in the present study, the volume of the drainage system was measured directly using a pressure/volume controller. As can be seen in Figure (1), the drainage system has two main parts: one connected to the top and the other connected to the bottom of the sample. Before performing the measurement, the drainage system was emptied using compressed air and then each part was connected to the pressure/volume controller keeping the valve of the drainage system closed. The connection pipe between the pressure/volume controller and the drainage system was filled with water. Then a small pressure was applied by the pressure/volume controller and its volume was set to zero before opening the valve of the drainage system. By opening the valve, the fluid flows into the drainage system. As soon as the first drop of the fluid flows out of the porous stone, the pressure/volume controller is stopped and the volume of the fluid which has filled the corresponding part of the drainage system is directly given by the volume change of the pressure/volume controller. The measurement was repeated for each part of the drainage system and the total volume of the drainage system was evaluated equal to $V_L = 2300\,\text{mm}^3$.

The compressibility of the drainage and pressure measurement systems $c_L$ is evaluated by applying a fluid pressure and by measuring the corresponding volume change in the pressure/volume controller. A metallic sample is installed inside the cell to prevent the fluid to go out from the drainage system. Fluid mass conservation is written in the following equation which is used to calculate the compressibility $c_L$ of the drainage system:

$$\frac{dV_L}{V_L} = \left(c_L + c_{fL}\right)dp_f \tag{29}$$

$dp_f$ and $dV_L$ are respectively the applied pore pressure and the volume change measured by the pressure/volume controller. For a single measurement, the volume change $dV_L$ accounts also for the compressibility of the pressure/volume controller and of the lines used to connect the pressure/volume controller to the main drainage system. To exclude the compressibility of these parts, a second measurement is done only on the pressure/volume controller and the connecting lines. The volume change $dV_L$ used in equation (29) is the difference between these two measurements. The measurements were performed separately for the two parts of the drainage system with volume $V_{L1}$ and $V_{L2}$ respectively. The compressibility of the entire system $c_L$ is simply obtained as the weighted average of the compressibilities of each part ($c_{L1}$ and $c_{L2}$ respectively):

$$c_L = \frac{V_{L1}}{V_L}c_{L1} + \frac{V_{L2}}{V_L}c_{L2} \tag{30}$$

The estimated value is $c_L = 0.117 \text{GPa}^{-1}$. This is equivalent to the compressibility that can be obtained in a single measurement in which the pressure/volume controller is connected simultaneously to the both parts of the drainage system using a T-connection.

The parameter $\beta$ and the thermal expansivity of the drainage system $\alpha_L$ are evaluated using the results of an undrained heating test performed using a metallic sample with the measurement of the fluid pressure change in the drainage system. For the metallic sample $\phi = 0$ and $c_d = c_s$ so that equation (24) is reduced to the following equation:

$$\Lambda^{mes} = \frac{\beta\left(\alpha_{fL} - \alpha_L\right)}{c_{fL} + c_L} \tag{31}$$

The thermal expansion coefficient $\alpha_{fL}$ and the compressibility $c_{fL}$ of water are known as functions of temperature and fluid pressure. As these variations are highly non-linear, the parameters $\beta$ and $\alpha_L$ cannot be evaluated directly but are back analysed from the calibration test results: the undrained heating test of the metallic sample is simulated analytically using equation (31) with a step by step increase of the temperature. For each step the corresponding water thermal expansion and compressibility are used [19]. The parameters $\beta$ and $\alpha_L$ are back-calculated by minimizing the error between the measurements and the computed results using a least-square algorithm. The test result and the back analysis are presented in Figure (2). The parameter $\beta$ is found equal to 0.46 and the thermal expansion coefficient of the drainage system $\alpha_L$ for this test is found equal to $1.57 \times 10^{-4} \, (°C)^{-1}$.

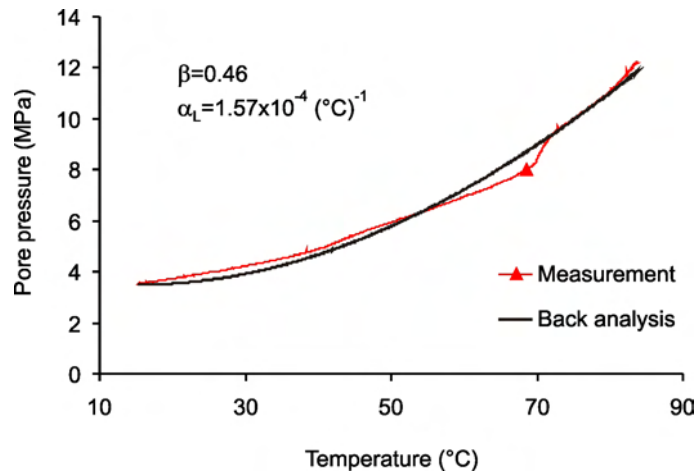

**Figure 2- Calibration test for the evaluation of the temperature ratio $\beta$ and the thermal expansion $\alpha_L$ of the drainage system-comparison.**

The evaluation of the compressibility $\kappa_L$ which represents the effect of the confining pressure on the volume of the drainage system is performed using an analytical method. As can be seen in Figure (1), only a part of the drainage system which is the pipe connected to the top of the sample, is influenced by the confining pressure. The effect of the confining pressure on the variations of the volume of this pipe can be evaluated using the elastic solution of the radial displacement of a hollow cylinder. Considering a hollow cylinder with inner radius $a$ and outer radius $b$ and radial stresses $p_i$ and $p_o$, respectively at the inner and outer boundaries, the well-known Lamé solution holds for the radial displacement $u(r)$ :

$$u(r) = Ar + \frac{B}{r} \tag{32}$$

The integration constants $A$ and $B$ are obtained from the boundary conditions:

$$A = \frac{p_i a^2 - p_o b^2}{2\left(\lambda + \mu\right)\left(b^2 - a^2\right)} \tag{33}$$

$$B = \frac{(p_i - p_o)a^2 b^2}{2\mu(b^2 - a^2)} \tag{34}$$

where $\lambda$ and $\mu$ are Lamé coefficients of the elastic cylinder.

The variation of the volume of the pipe with the inner radius $a$ and the length $L$ is given by:

$$dV_L = (2\pi a L) da \tag{35}$$

From equations (32) to (35), and by setting $p_i = 0$, $p_o = d\sigma$ the following expression is obtained for the variation of the volume of the drainage system under the effect of the confining pressure:

$$dV_L = -\frac{\pi a^2 b^2 L}{(b^2 - a^2)} \frac{\lambda + 2\mu}{\mu(\lambda + \mu)} d\sigma \tag{36}$$

Inserting equation (36) in equation (20) the following expression is obtained for the compressibility $\kappa_L$:

$$\kappa_L = \frac{\pi a^2 b^2 L}{(b^2 - a^2)V_L} \frac{\lambda + 2\mu}{\mu(\lambda + \mu)} \tag{37}$$

Equation (37) can be re-written using the Young's modulus $E$ and the Poisson's ratio $\nu$ ($\lambda = \frac{E\nu}{(1 + \nu)(1 - 2\nu)}$, $\mu = \frac{E}{2(1 + \nu)}$):

$$\kappa_L = \frac{4\pi a^2 b^2 L(1 - \nu^2)}{(b^2 - a^2)V_L E} \tag{38}$$

For the considered pipe $a = 0.25\,\text{mm}$, $b = 0.8\,\text{mm}$, $L = 900\,\text{mm}$, $E = 190\,\text{GPa}$ and $\nu = 0.3$. Inserting these values and $V_L = 2300\,\text{mm}^3$ in equation (38) we obtain $\kappa_L = 1.6 \times 10^{-3}\,\text{GPa}^{-1}$. This value is very small as compared to the compressibility $c_L = 0.117\,\text{GPa}^{-1}$ which takes into account the effect of the pore pressure variations on the volume of the drainage system. This is due to the fact that only a small part of the drainage system, less than 8% of its volume, is influenced by the confining pressure.

## 6   Examples of application of the correction method

Two examples of the application of the proposed correction method are presented for an isothermal undrained isotropic compression test and an undrained heating test performed on a fluid-saturated granular rock, Rothbach sandstone which has a porosity of 16% and is composed of 85% quartz, 12% feldspars and 3% clay. As can be seen in equations (25), (26), (27) and (28), the application of the correction method needs the knowledge of the drained and the unjacketed thermo-poro-elastic parameters of the tested material. The required parameters are evaluated and presented by Ghabezloo and Sulem [4]. The unjacketed modulus $K_s = 1/c_s$ of the rock was evaluated in an unjacketed compression test and found equal to 41.6 GPa. A drained isotropic compression test was performed with a loading-unloading cycle and the tangent drained bulk modulus $K_d = 1/c_d$ of the rock was found to be stress-dependent and could be approximated by the following expression:

$$\begin{array}{ll} K_d = 0.96\,\sigma_d + 0.70 & \sigma_d \le 9\,\text{MPa} \\[6pt] K_d = 0.07\,\sigma_d + 8.72 & \sigma_d > 9\,\text{MPa} \end{array} \qquad (K_d : \text{GPa}, \ \sigma_d : \text{MPa}) \tag{39}$$

The drained thermal expansion coefficient of the rock was evaluated in a drained heating test and found to be constant, equal to $28 \times 10^{-6}\,(^\circ\text{C})^{-1}$. For the correction of the results of the undrained heating test we assume that $\alpha_\phi = \alpha_d$.

## 6.1  Undrained isotropic compression test

An undrained isotropic compression test with a loading-unloading cycle is performed at ambient temperature. The initial confining pressure is 2.0 MPa and the initial pore pressure is 0.5 MPa. The confining pressure is increased to 12.0 MPa and then decreased to its initial value with a loading rate of 0.4 MPa/min. The initial temperature of the sample was 21.6°C. During the loading and unloading phases the temperature was increased to 22.6°C and then decreased to 21.2°C due to the quasi-adiabatic heating and cooling of the confining oil with a rate of about 0.1°C/MPa. This rate of heating/cooling could be decreased by using a slower loading rate. In an other experiment performed on a hardened cement paste [16], with a loading rate of 0.1 MPa/min (i.e. four times slower than the one used in this study), the resulted heating/cooling rate was reduced to 0.04°C/MPa. The effects of the temperature change are here sufficiently small to be neglected in this analysis of the test results. However, one should keep in mind that these effects could be important if the loading rate is much higher and if the thermal pressurization coefficient of the tested material is high. The volumetric strain response of the sample is presented in Figure (3) and shows the presence of a small quantity of non-elastic strains at the end of unloading phase. The variations of pore pressure with the confining pressure are presented in Figure (4). The elastic tangent undrained bulk modulus $K_u = 1/c_u$ and the elastic tangent Skempton coefficient $B$ can be evaluated as the slope of the unloading parts of the curves in Figures (3) and (4). The evaluated parameters are presented in Figure (5) as a function of Terzaghi effective stress. This figure shows the increase of the undrained bulk modulus and the decrease of the Skempton coefficient with the increase of the Terzaghi effective stress, which is compatible with the stress dependency of the drained bulk modulus of the tested rock, as can be seen in equation (39). As the test was performed at ambient temperature, for the correction of the evaluated undrained bulk modulus and Skempton coefficient using equations (25) and (26) we take $\rho_{fL} = \rho_f$ and $c_{fL} = c_f$. For the correction of the tangent moduli, the relevant tangent drained bulk modulus is calculated at each point using equation (39) as a function of Terzaghi effective stress corresponding to that point. The corrected parameters are presented in Figure (5) and can be compared with the measured values. We can see that the corrected values of both parameters, the undrained bulk modulus and the Skempton coefficient, are greater than the measured ones.

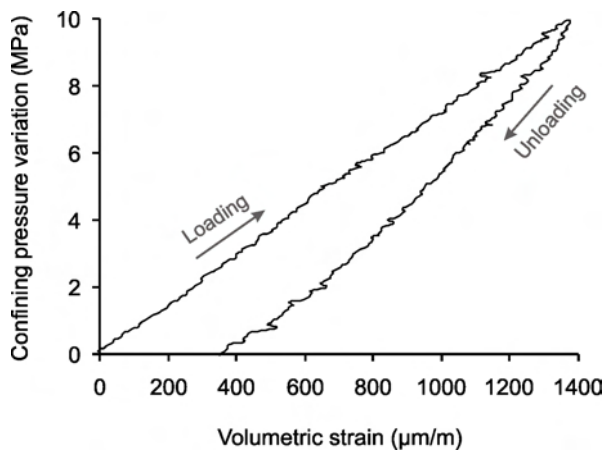

**Figure 3- Undrained isotropic compression test: volumetric strain response**

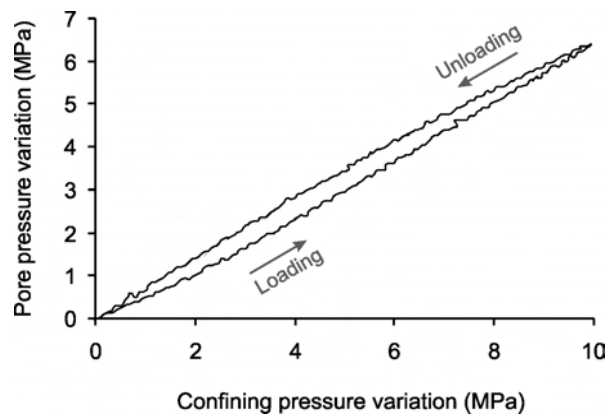

**Figure 4- Undrained isotropic compression test: pore pressure response**

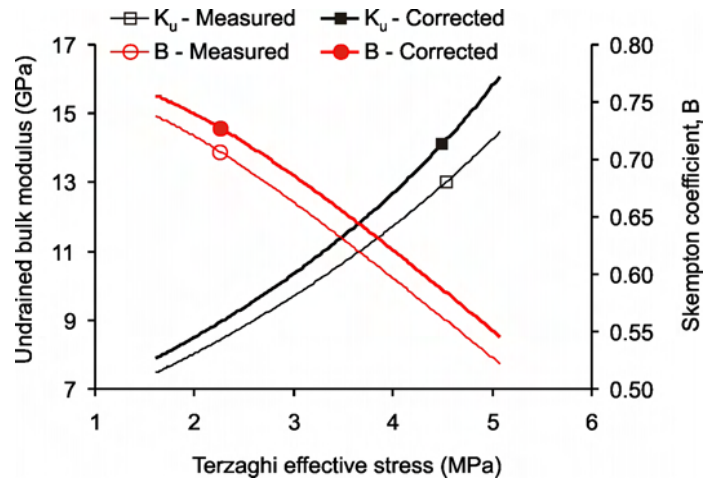

**Figure 5- Variations of the measured and corrected elastic tangent undrained bulk modulus and Skempton coefficient with Terzaghi effective stress**

## 6.2 Undrained heating test

The performed undrained heating test is presented in Ghabezloo and Sulem [4]. The undrained heating test was performed under constant isotropic stress equal to 10 MPa. The initial temperature of the sample was 20°C and the rate of temperature change was 0.2°C/min. The measured pore pressure during the test is presented in Figure (6) as a function of the temperature increase. One can observe the non-linear increase of the pore pressure with the temperature up to the state for which the pore pressure in the sample reaches the confining pressure. At this point the pore fluid of the sample infiltrates between the sample and the rubber membrane so that the pore fluid pressurization is stopped. The slope of pore pressure curve versus the temperature gives the thermal pressurization coefficient $\Lambda$ and is presented in Figure (7) as a function of the temperature increase. The non-linearity of the observed thermal pressurization coefficient is due to the (effective) stress-dependent compressibility of the sandstone and also to the temperature and pressure dependent compressibility and thermal expansion of the pore water. More details about the mechanism governing this non-linear behaviour can be found in Ghabezloo and Sulem [4]. At the beginning of the test the volumetric strain could not be recorded due to a failure of the displacement sensors inside the triaxial cell. Therefore the volumetric strain-temperature curve, presented in Figure (8), only starts from 40°C. The slope of this curve gives the undrained thermal expansion coefficient and is presented in Figure (9) as a function of the temperature increase. We can see the increase of this coefficient with temperature which is mostly due to the significant increase of the thermal expansion coefficient of the water with temperature.

The test results are corrected using equations (27) and (28) for the effect of the dead volume of the drainage system. At each data point the relevant thermal expansion and compressibility of water are used as a function of the current pore pressure and temperature [19]. The drained compressibility is also calculated at each point as a function of Terzaghi effective stress (equation (39)). The variation of the mechanical properties of the sandstone with temperature is neglected in the analysis. It should be mentioned that the very good compatibility obtained by Ghabezloo and Sulem [4] between the results of an analytical simulation of this test and the experimental results showed the negligible effect of this assumption. The corrected values of the pore pressure, thermal pressurization coefficient, volumetric strain and undrained thermal expansion coefficient are presented along with the measured values respectively on Figures (6), (7), (8) and (9). The corrected pore pressure and thermal pressurization coefficient are more important that the measured values. The corrected volumetric strain and undrained thermal expansion coefficient are slightly smaller than the measured values, but the average difference between the corrected and measured curves is about 1%. As can be seen in equations (27) and (28), apart from the properties of the drainage system, the correction depends also on the measured values of thermal pressurization coefficient and undrained thermal expansion coefficient. For the same triaxial cell, the correction of the observed pore pressure during an undrained heating test performed on a hardened cement paste [5] is more important than the one obtained here for Rothbach sandstone.

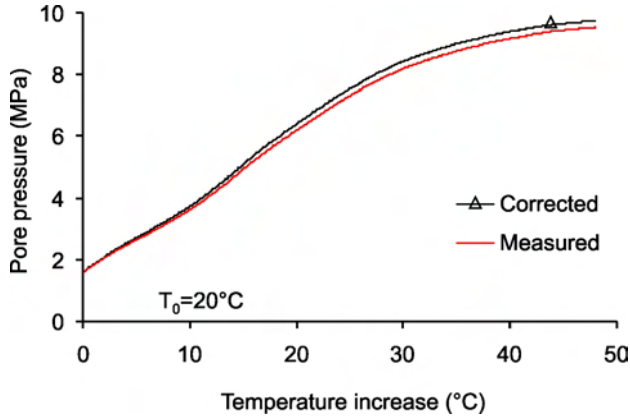

**Figure 6- Undrained heating test: measured and corrected pore pressure response**

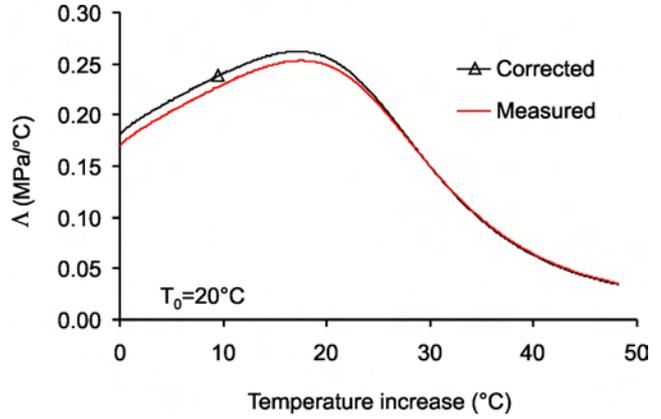

**Figure 7- Undrained heating test: variations of the measured and the corrected thermal pressurization coefficient with temperature**

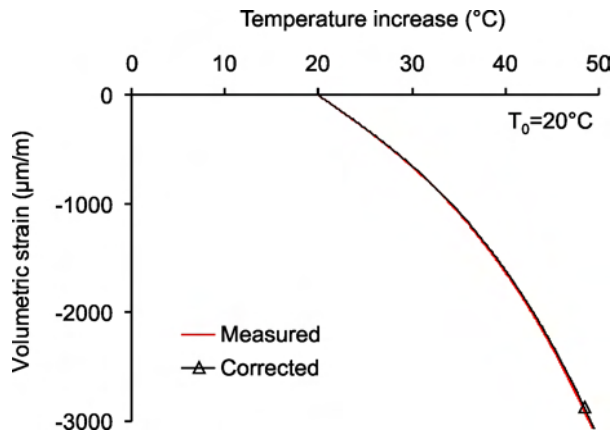

**Figure 8- Undrained heating test: measured and corrected volumetric strain response**

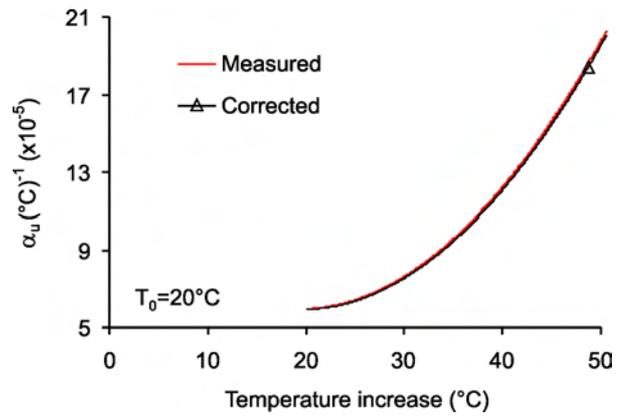

**Figure 9- Undrained heating test: variations of the measured and the corrected undrained thermal expansion coefficient with temperature**

# 7   Parametric study

In this section, a parametric study on the error made on the measurement of different undrained thermo-poro-elastic parameters is presented. The error on a quantity $Q$ is evaluated as $(Q_{\text{measured}} - Q_{\text{real}})/Q_{\text{real}}$ and takes positive or negative values with indicates if the measurement overestimates or underestimates the considered quantity.

As shown in the following, among the different parameters appearing in equations (25) to (28), the porosity $\phi$ of the tested material, its drained compressibility $c_d$ and the ratio of the volume of the drainage system to the one of the tested sample, $V_L/V$ are the most influent parameters. For this parametric study the parameters of the drainage system are taken equal to the ones of the triaxial system used in this study, as presented in section 4. We take also $c_s = c_\phi = 0.02\text{GPa}^{-1}$ and $\alpha_\phi = 3\times10^{-4}\left(°C\right)^{-1}$ which are typical values. Figure (10) presents the error on the measurement of the Skempton coefficient $B$ as a function of the sample porosity $\phi$, for three different values of drained compressibility and two different values of the ratio $V_L/V$. Three different values of the drained compressibility are considered, respectively equal to $0.03\text{GPa}^{-1}$, $0.1\text{GPa}^{-1}$ and $0.5\text{GPa}^{-1}$, which covers a range from a rock with a low compressibility to a relatively highly compressible rock. The porosity is varied from 0.05 to 0.35. The ratio $V_L/V$ is taken equal to 0.025 which corresponds to the conditions of the triaxial system used in this study. We analyze also the effect of a greater volume of the drainage system on the measurement errors by choosing another value twice bigger, equal to 0.05. We can see in Figure (10) that the error on the measurement of the Skempton coefficient is always negative (the measurement underestimates the real value, see also equation (25)) and covers an important range between 2% and 50%. Considering the tested material, the measurement error is more significant for low-porosity rocks with low-compressibility. We can also see the significant effect of the volume of the drainage system on the measurement error. The error of the measurement for the undrained compressibility $c_u$ is presented in Figure (11) where we can observe that it is more important for low-porosity rocks and for greater volume of the drainage system. As opposite to what we observed for the Skempton coefficient, the error of the measurement of $c_u$ is more important when the tested material is more compressible.

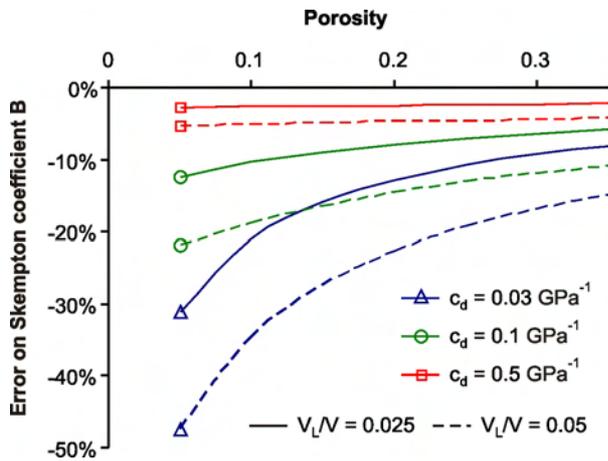

**Figure 10- Parametric study of the error on Skempton coefficient $B$**

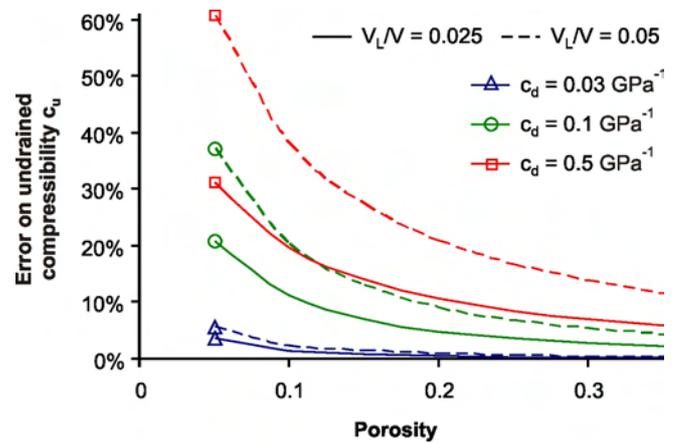

**Figure 11- Parametric study of the error on the undrained compressibility $c_u$**

Figures (12) and (13) show the errors corresponding to the measurements of the thermal pressurization coefficient $\Lambda$ and of the undrained thermal expansion coefficient $\alpha_u$ respectively. We observe that the error of the measurement for $\Lambda$ varies between -40% and +10%, which shows that the measured value may be smaller or greater than the real one. As for the isothermal undrained parameters, the error is more important for low-porosity materials and for a greater volume of the drainage system. The error for the undrained thermal expansion coefficient $\alpha_u$ varies between -6% and +4%, which is a narrower range, as compared to the other undrained parameters.

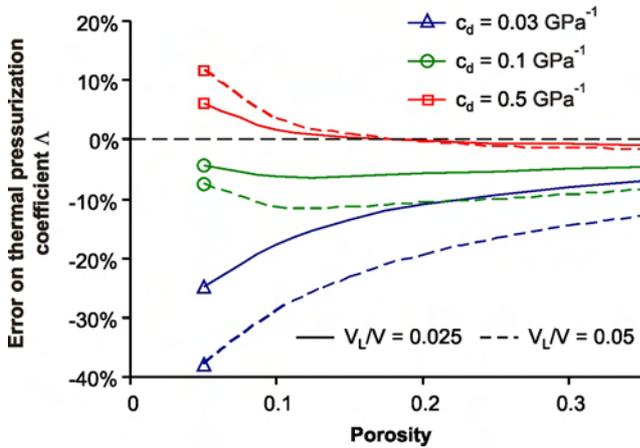

**Figure 12- Parametric study of the error on the thermal pressurization coefficient** $\Lambda$

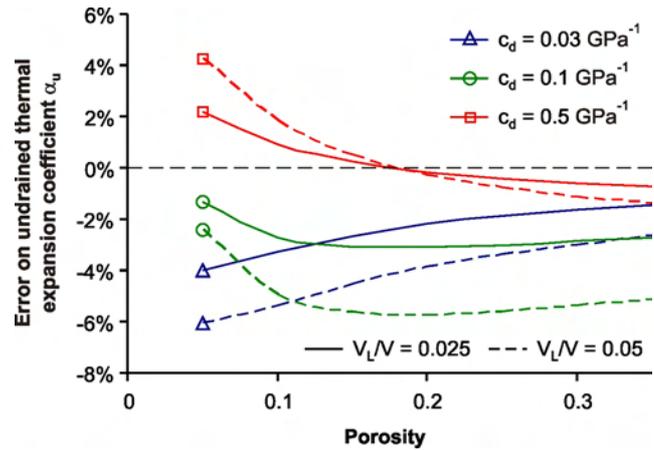

**Figure 13- Parametric study of the error on the undrained thermal expansion coefficient** $\alpha_u$

## 8 Conclusions

For evaluation of the undrained thermo-poro-elastic properties of saturated porous materials in conventional triaxial cells, it is important to take into account the effect of the dead volume of the drainage system. The compressibility and the thermal expansion of the drainage system along with the dead volume of the fluid filling this system, influence the measured pore pressure and volumetric strain during an undrained thermal or mechanical loading in a triaxial cell. The correction methods previously presented by Wissa [1], Bishop [2] and Ghabezloo and Sulem [4] only permit to correct the measured pore pressures during an undrained isotropic compression test or an undrained heating test. An extension of these methods is presented in this paper to correct also the measured volumetric strain and consequently the measured undrained bulk compressibility and undrained thermal expansion coefficients during these tests. Two examples of application of the proposed correction method are presented on the results of an undrained isotropic compression test and an undrained heating test performed on a fluid-saturated granular rock. A parametric study has demonstrated that the porosity $\phi$ of the tested material, its drained compressibility $c_d$ and the ratio of the volume of the drainage system to the one of the tested sample, $V_L/V$ are the key parameters which influence the most the error induced on the measurements by the drainage system. It has also shown that the Skempton coefficient, the thermal pressurization coefficient and the undrained compressibility measurements are much more affected than the measurement of the undrained thermal expansion coefficient.

## 9 Acknowledgment

The authors wish to thank François Martineau for his assistance in the experimental work.

## 10 References

[1] Wissa AE, Pore pressure measurement in saturated stiff soils. ASCE Journal of Soil Mechanics and Foundations Div, 1969;95(SM 4):1063-1073.

[2] Bishop AW, Influence of system compressibility on observed pore pressure response to an undrained change in stress in saturated rock. Geotechnique, 1976;26(2):371-375.

[3] Mesri G, Adachi K, Ullrich CR, Pore-pressure response in rock to undrained change in all-round stress. Geotechnique, 1976;26(2):317-330.

[4] Ghabezloo S, Sulem J, Stress dependent thermal pressurization of a fluid-saturated rock. Rock Mechanics and Rock Engineering, 2009;42(1):1-24.

[5] Ghabezloo S, Sulem J, Saint-Marc J, The effect of undrained heating on a fluid-saturated hardened cement paste. Cement and Concrete Research, 2009;39(1):54-64.

[6] Berryman JG, Effective stress for transport-properties of inhomogeneous porous rock. Journal of Geophysical Research-Solid Earth, 1992;97(B12):17409-17424.

[7] Hill R, The elastic behaviour of a crystalline aggregate. Proceedings of the Physical Society. Section A, 1952;5:349.

[8] Berge RA, Pore compressibility in rocks. Poromechanics: A Tribute to Maurice A. Biot, Proceedings of the Biot Conference on Poromechanics, Louvain-la Neuve, Belgium, 351, 1998.

[9] Berge PA, Berryman JG, Realizability of negative pore compressibility in poroelastic composites. Journal of Applied Mechanics-Transactions of the Asme, 1995;62(4):1053-1062.

[10] Brown RJS, Korringa J, On the dependence of the elastic properties of a porous rock on the compressibility of the pore fluid. Geophysics, 1975;40(4):608-616.

[11] Zimmerman RW, Somerton WH, King MS, Compressibility of porous rocks. Journal of Geophysical Research-Solid Earth and Planets, 1986;91(B12):2765-2777.

[12] Skempton AW, The pore pressure coefficients a and b. Geotechnique, 1954;4:143-147.

[13] Gassmann F, Über die elastizität poroser medien. Veirteljahrsschrift der Naturforschenden Gesellschaft in Zürich, 1951;96:1-23.

[14] Palciauskas VV, Domenico PA, Characterization of drained and undrained response of thermally loaded repository rocks. Water Resources Research, 1982;18(2):281-290.

[15] McTigue DF, Thermoelastic response of fluid-saturated porous rock. Journal of Geophysical Research-Solid Earth and Planets, 1986;91(B9):9533-9542.

[16] Ghabezloo S, Sulem J, Guedon S, Martineau F, Saint-Marc J, Poromechanical behaviour of hardened cement paste under isotropic loading. Cement and Concrete Research, 2008;38(12):1424-1437.

[17] Lockner DA, Stanchits SA, Undrained poroelastic response of sandstones to deviatoric stress change. J. Geophys. Res, 2002;107:2353.

[18] Sulem J, Ouffroukh H, Hydromechanical behaviour of fontainebleau sandstone. Rock Mechanics and Rock Engineering, 2006;39(3):185-213.

[19] Spang B. Excel Add-In for Properties of Water and Steam in SI-Units. http://www.cheresources.com/iapwsif97.shtml, 2002.